\documentclass[]{interact}   

\usepackage{amsmath,amsfonts,amssymb}						
\usepackage{nicefrac} 										
\usepackage{mhsetup} 										
\usepackage{mathtools}										

\usepackage{siunitx} 										
\usepackage{xcolor}											

\usepackage{tabularx}
\newcolumntype{Y}{>{\centering\arraybackslash}X}		

\DeclareSIUnit\kayser{\per \cm}

\newcommand{\bands}[2]{\mbox{(#1--#2)}}		

\newcommand{\dun}[2]{\ensuremath{\mathrm{Y}_{#1#2}}}

\newcommand{\dunG}[2]{\ensuremath{\gamma_{#1#2}} }

\newcommand{\fone}{$\mathrm{F}_1$}
\newcommand{\ftwo}{$\mathrm{F}_2$}
\newcommand{\Xstate}{$X(1)^2\Sigma^+$}
\newcommand{\bstate}{$(2)^2\Sigma^+$}
\newcommand{\astate}{$(1)^2\Pi$}
\newcommand{\dSig}{$^2\Sigma^+$}
\newcommand{\dPi}{$^2\Pi$}
\newcommand{\dPione}{\ensuremath{^2\Pi_{1/2}}}
\newcommand{\dPithree}{\ensuremath{^2\Pi_{3/2}}}

\newcommand{\wn}{\mbox{cm$^{-1}$}}

\begin{document}
	
	\title{The electronic system \bstate\ and \astate\ of LiCa}  
	
	\date{\today}
	\author{
	\name{Julia Gerschmann\textsuperscript{a,b} and Erik Schwanke\textsuperscript{a,b} and Silke Ospelkaus\textsuperscript{a,b} and Eberhard Tiemann\textsuperscript{a,b}\thanks{CONTACT {\email{silke.ospelkaus@iqo.uni-hannover.de  or  tiemann@iqo.uni-hannover.de}} {ORCID Tiemann 0000-0002-4690-7738, Schwanke 0000-0002-9607-0753}}}
	\affil{ \textsuperscript{a} Institute of Quantum Optics, Leibniz Universit\"at Hannover, Welfengarten 1, 30167 Hannover, Germany}
	\affil{\textsuperscript{b} Laboratory for Nano- and Quantum Engineering,  Leibniz Universit\"at Hannover, Schneiderberg 39, 30167 Hannover, Germany}
}	
	\maketitle

	\begin{abstract}
		High resolution Fourier transform spectroscopy and Laser induced fluorescence has been performed on LiCa in the infrared spectral range. We analyze  rovibrational transitions of the \bstate--\Xstate\ system of LiCa and find  the \bstate\ state to be perturbed by spin-orbit coupling to the \astate\ state. We study the coupled system obtaining molecular parameters for the \bstate\ and  the \astate\ state together with effective spin-orbit and spin-rotation coupling constants. The coupled system has also been evaluated by applying a potential function instead of rovibrational molecular parameters for the state \bstate. An improved analytic potential function of the \Xstate\ state is derived, due to the extension of the observed rotational ladder.
	\end{abstract}

	\section{Introduction}
	
In the area of ultracold molecules, the interest in alkali-alkaline earth diatomic molecules has increased during recent years, since their ground state $^2\Sigma^+$  has a magnetic dipole moment from the almost free electron spin in addition to the electric one, see theoretical estimation of 0.437 a.u. $\approx$ 1.11 Debye for the nonrotating molecule in the vibrational ground state \cite{kotochigova_ab_2011}. We mention only few examples, like the production of degenerated Bose-Fermi mixtures of Li and Sr \cite{Zhu:20} and the observation of Feshbach resonances in cold collisions of Rb and Sr \cite{barbe_observation_2018}. In several studies Yb is replacing the alkaline earth part, e.g. in LiYb to study quantum degenerated mixtures \cite{Hara:11} or Feshbach resonances \cite{Green:20}.

There are several ab initio calculations for this class of molecules, for example \cite{pototschnig_electric_2016,pototschnig_vibronic_2017,gopakumar_ab_2013,gopakumar_dipole_2014}, giving good guide lines for spectroscopic studies which in turn will help to extend the research on ultracold ensembles with such atomic mixtures and/or molecules. The ground state \Xstate\ of these molecules correlates with the atom pair ground state $^2S_{1/2}$+$^1S_0$, which leads to a single molecular state energetically well separated from the excited states. The two first excited atom pair asymptotes $^2P_{1/2,3/2}$+$^1S_0$ and $^2S_{1/2}$+$^3P_{0,1,2}$ which could be energetically fairly close to each other, lead according to ab initio calculations to families of doublet and quartet molecular states with $\Sigma^+$ and $\Pi$ character. In our present case of LiCa the two lowest states \bstate\ and \astate\ are predicted to be sufficiently separated from the other electronic states justifying to treat them as a coupled pair of molecular states neglecting all other states. The state \bstate\ is embedded in state \astate, thus perturbations are expected for the band spectra of the \bstate\ - \Xstate\ system lying in the near infrared range and being significantly stronger than the \astate\ - \Xstate\ system in the mid infrared range.

In our research group, the molecules LiCa \cite{ivanova_x_2011,stein_spectroscopic_2013}, LiSr \cite{schwanke_laser_2017,schwanke_2020} and KCa \cite{gerschmann_laser_2017} were generated in a heat pipe oven and investigated with high resolution spectroscopy using laser-induced fluorescence (LIF). 
For the LiCa molecule, the analytical potential and the Dunham description of the ground state \Xstate\ (up to v''= 19) were reported in ref. \cite{ivanova_x_2011}, applying extended fluorescence progressions after laser excitations of the electronic system \Xstate\ - $(4)^2\Sigma^+$ in the visible spectral range. In ref. \cite{stein_spectroscopic_2013} the potentials and Dunham coefficients for the excited states \bstate\ and $(4)^2\Sigma^+$ were derived. From the band spectra \bstate\ - \Xstate\ vibration levels of v'= 0 and 1 and only 2 lines for v'= 2 of the excited state were identified using the recorded thermal emission and few LIF spectra. Surprisingly, no perturbation was detected in that range.

The present work describes a significantly extended investigation of the \bstate\ state of LiCa and provides a new recording of the thermal emission with a higher resolution than in ref.  \cite{stein_spectroscopic_2013}  and better signal-to-noise ratio. Additionally, numerous LIF experiments are performed for obtaining an unambiguous assignment of the rotational quantum numbers. With the expansion of the rovibrational quantum numbers of the \bstate\ state, the range of the derivable analytical potential is enlarged. By the detailed study, local perturbations for all vibrational states are detected, which arise due to spin-orbit interactions between the states \bstate\ and \astate\, as already reported for LiSr  \cite{schwanke_2020}. This leads to important information on the \astate\ state and to the experimental determination of the sign of the spin-rotation splitting, not derivable from the data in ref.  \cite{ivanova_x_2011}. 	
	
\section{Experiment and quantum number assignment}
	\label{exp.}
We use the procedure for preparing LiCa samples in a heat pipe described in ref.  \cite{ivanova_x_2011}. The metals are loaded into a 3-section heat pipe (as for KCa  \cite{gerschmann_laser_2017}), whereby the different temperatures of the sections generate similar vapor pressures for the metals. It is also important to place the metals separately so that they only mix in the gas phase to form molecules. The heat pipe consists of a 88 cm long tube (steel type 1.4841 ) with an inner diameter of 3 cm, the middle section is located in the center of the oven. The  tube is covered inside with a steel mesh so that the condensed metal can move back into the heated section. The ends of the tube are closed with wedged windows from BK7. The areas directly at the windows are water cooled. 

10 g of Ca are placed in the hottest section (approx. \SI{950}{\celsius}) of the heat pipe. The section with 9 g of Li is kept at \SI{620}{\celsius}-\SI{660}{\celsius}. The other side with Ca, facing the spectrometer, is held at \SI{780}{\celsius}-\SI{790}{\celsius}. The thermal emission is recorded with a Fourier transform spectrometer (IFS 120 HR, Bruker) set at a resolution of 0.02 \si{\kayser}. The spectrum is provided in ASCII format as supplementary material \cite{sup}. 
Fluorescence is generated with a single mode diode laser stabilized to a wavemeter, and the LIF spectra are recorded with a resolution of 0.05 \si{\kayser} by the FT spectrometer simply to save time. 

	\begin{figure*}[htb]
	\includegraphics[width = 0.9\textwidth]{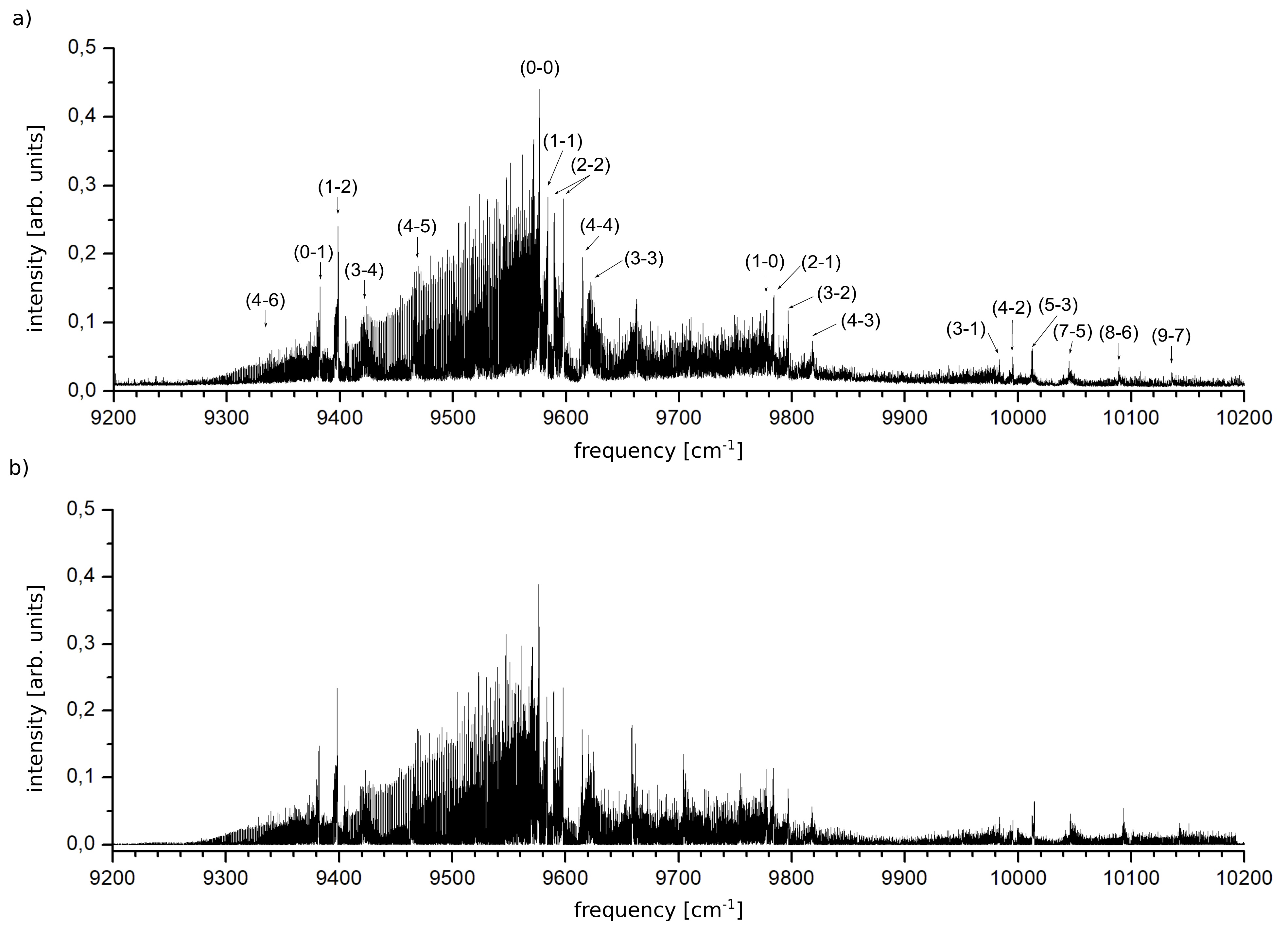}    
	\caption{\label{fig:lica_sim}
		a) Recorded spectrum of LiCa with assigned vibrational bands; b) simulated spectrum, perturbations are considered for $v'<5$.}
\end{figure*}

	\begin{figure*}[htb]
	\includegraphics[width = 0.9\textwidth]{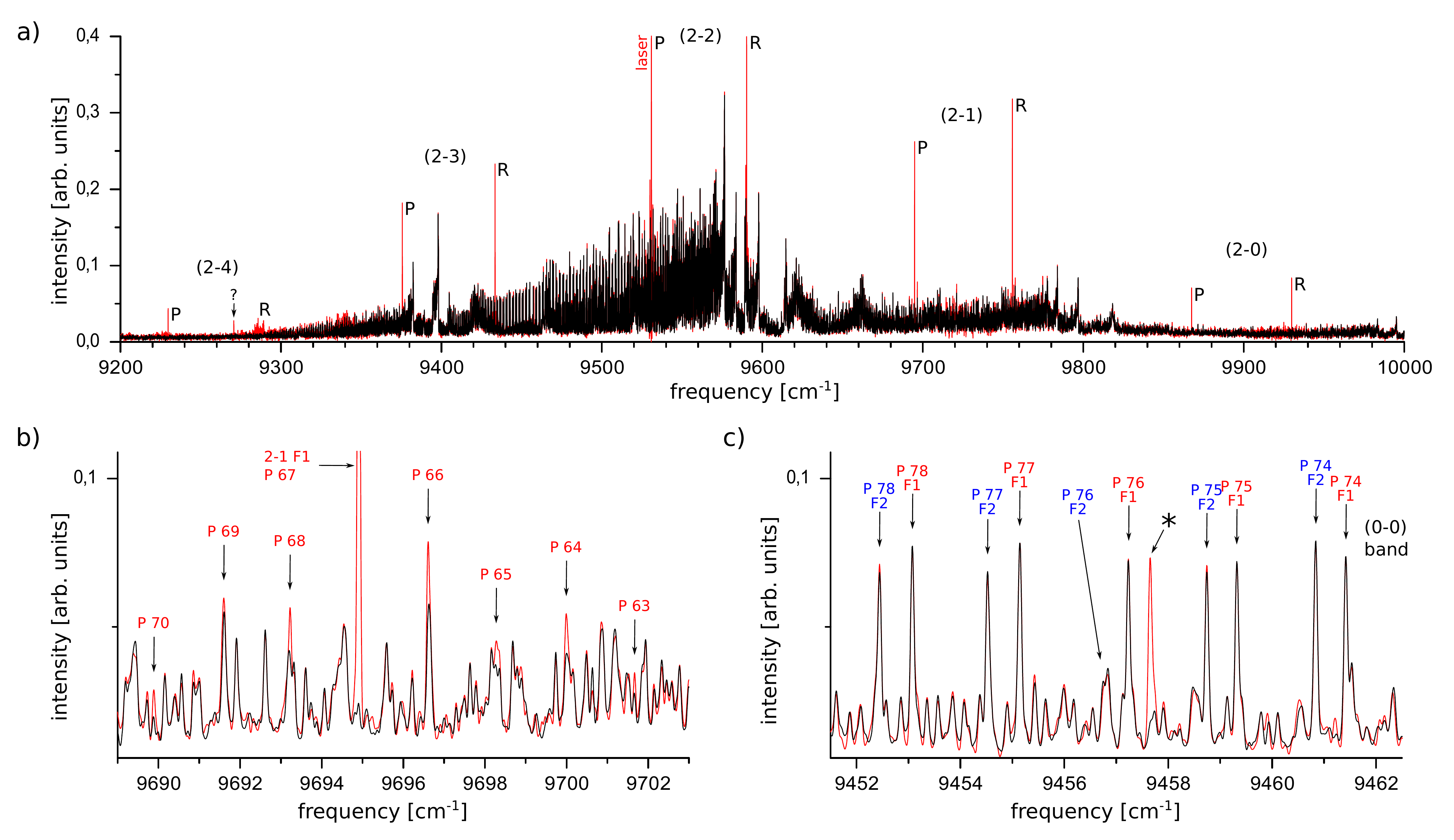}
	\caption{\label{fig:lica_lif}
		Fluorescence lines (red) overlapped with the thermal emission spectrum (black). Excitation is at 9530.8435 \si{\kayser} in the \bands{2}{2} band. a) whole spectrum with P-R pairs of bands (2-v") v"=0 to 4; b) collisional satellites from excitation of P67 F$_1$ (2-1); c) \bands{0}{0} band region after excitation by the P67 F$_1$ (2-2) transition of the main isotopologue, the only fluorescence line marked by an asterisk, is identified as the P70 F$_2$ (0-0) line of the isotopologue $^6$Li$^{40}$Ca. The weak features are not noise, see note in section \ref{outlook}.}
\end{figure*}

The thermal emission spectrum of LiCa is shown in Fig. \ref{fig:lica_sim}. Compared to the spectrum reported in ref.  \cite{stein_spectroscopic_2013}, there is almost no background intensity and thus the signal-to-noise ratio is significantly improved. One can see several band heads, which are mostly red shaded. Zooming into the recording, many lines of the \bands{0}{0} band can be clearly distinguished from the lines of other bands and assigned using only the thermal emission spectrum (see Fig. \ref{fig:lica_lif} c)), as well as some lines of the \bands{1}{1} band with low rotational quantum numbers. This procedure was applied in  \cite{stein_spectroscopic_2013}  adding few LIF experiments to confirm the assignment. 

In the present study, many laser excitations in the system \bstate\ - \Xstate\ made it possible to identify lines up to $v'=4$ in the excited state. With the help of a calibrated wavemeter the laser could be set with an accuracy of 0.0005 \si{\kayser}. Fig. \ref{fig:lica_lif} gives examples of fluorescence detection, which appears as enhanced intensity (red) compared to the thermal emission (black). Fig. \ref{fig:lica_lif} a) shows a progression of fluorescence lines after the excitation of a P line of the (2-2) band at 9530.8435 \si{\kayser}. Since the ground state is known in great detail from ref.  \cite{ivanova_x_2011}, the rotational quantum numbers can be determined with high reliability based on the P-R spacings in several bands, even if the lines are perturbed and thereby shifted. Collisional satellites (Fig. \ref{fig:lica_lif} b)) were often observed in fluorescence spectra. These are very helpful when assigning the lines to a common F-component\footnote[1]{spin-rotation splitting, see section \ref{theory}}, especially in perturbed areas. The F component cannot be recognized from the P-R spacings, because the difference is below the experimental resolution of 0.02 \si{\kayser}. Several lines belonging to the isotopologue $^6$Li$^{40}$Ca, which has only an abundance of 7.3\% compared to 89.7\% of the main isotopologue $^7$Li$^{40}$Ca, were excited by chance in the LIF experiments. The example in Fig. \ref{fig:lica_lif} c) shows the fluorescence recording (red spectrum) with a single fluorescence enhancement that matches the P70 F$_2$ of the \bands{0}{0} band of $^6$Li$^{40}$Ca. The laser frequency is primarily set for exciting P67 F$_1$ in the (2-2) band of the main isotope but overlaps with the associated R68 F$_2$ (0-0) line of $^6$Li$^{40}$Ca. It can also be seen in Fig. \ref{fig:lica_lif} c) that line P76 F$_2$, expected on the left of F$_1$ is missing or is weak in the regular series of assigned thermal emission lines because it may be shifted and has a low intensity, both due to perturbation.  This line could be identified by the final simulation, see section \ref{sec:resdis}, and therefore marked in the figure.
	
The spectra from ref.  \cite{ivanova_x_2011}  and their assignment of spin-rotation splitting are the starting point of the analysis of the present extended data set, and we perform first fits of Dunham coefficients and an analytic potential for the \bstate\ state (see definition in section \ref{theory}) without taking into account the interaction with the \astate\ state. For the next iteration, a simulation of the thermal spectrum was calculated with the fitted potential and the ground state potential from ref.  \cite{ivanova_x_2011}  and repeatedly compared with the recorded spectrum to obtain more assigned lines for the next fit iteration. Spectral ranges, that were not yet well described or are perturbed, could be identified in the simulated spectrum in order to investigate them in LIF experiments. Their results were included in the next iteration step. In this way vibrational levels from v'=0 to 4 could be undoubtedly assigned. Lines with higher v' were too strongly overlapped or too weak to obtain a reliable assignment even for cases where the band head is seen in Fig. \ref{fig:lica_sim} a). In total from the laser excitations, the resulting progressions and satellite lines and the detailed study of the thermal emission spectrum, 758 different levels of the excited state \bstate\ were derived by adding the ground state level energy to the measured transition energy. The LiCa spectrum simulation using the final results of the analysis including the perturbation data of section \ref{sec:resdis} is pictured in Fig. \ref{fig:lica_sim} b) and shows a very convincing agreement with the recorded spectrum (Fig. \ref{fig:lica_sim} a)). 

\begin{figure*}[htb]
	\includegraphics[width = \textwidth]{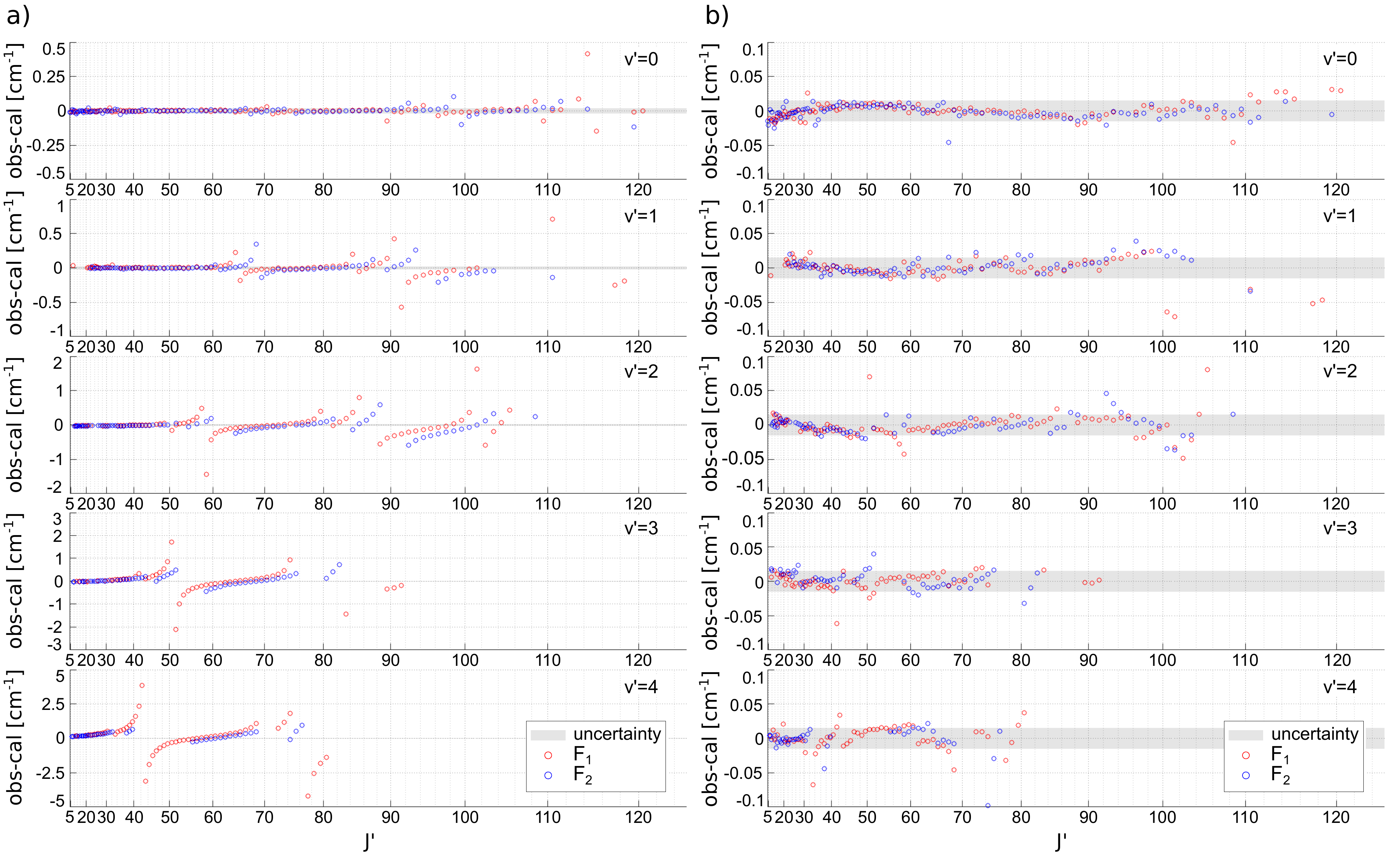}  
	
	\caption{\label{fig:lica_obs-cal}
		Difference between observed and calculated transition frequencies without considering the coupling (a) and after the deperturbation (b). Note the enlarged energy scale in part (b) to demonstrate the quality of deperturbation. The shaded area indicates the experimental uncertainty.}
\end{figure*}

The results of this first analysis are summarized in Fig. \ref{fig:lica_obs-cal} a). In this figure, we plot as function of the total angular momentum J' the differences between the experimentally derived  and the calculated level energies of \bstate\ with the fitted potential of \bstate\ including the spin-rotation energy. Systematic deviations can be recognized which should be caused by spin-orbit interactions with the \astate\ state around the crossings of the rovibrational ladders of the two states. The detailed discussion of the deviations and these interactions will be taken into account in the so-called deperturbation procedure to derive the molecular parameters for the uncoupled states and their coupling magnitudes. The next section will set up the appropriate physical model. The line assignment is provided in ASCII format as supplementary material \cite{sup}.

\section{The coupled system \bstate\ - \astate\ }
		\label{theory}
		
The observed spectrum belongs to the transition \bstate\ - \Xstate. Assuming a small influence of the coupling to \astate\ according to our earlier report on LiCa  \cite{stein_spectroscopic_2013}  we applied in section \ref{exp.} as a first approach the conventional Dunham representation of the molecular levels $|^2\Sigma ^+,v,J,e/f>$ with vibrational quantum number v and total angular momentum J as given in the following definition:
	\begin{equation}%
		\label{eq:DunhamExpansionaSig}%
		E_\mathrm{Dun}^{\text{(a)}}(v,J, \nicefrac{e}{f}) = \sum_{m,n} \mathrm{Y}^{\text{(a)}}_{m,n}[v+\nicefrac{1}{2}]^m[(J\mp\nicefrac{1}{2})(J\mp\nicefrac{1}{2}+1)]^n.
	\end{equation}%
Here, the upper sign holds for $e$ (\fone) levels and the lower sign for $f$ (\ftwo) levels. This is the representation in Hund's coupling case (a). The basis vector in Hund's case (b) will be defined by $|^2\Sigma ^+,v,(NS)J>$ with rotational quantum number N and the electronic spin S resulting to $\vec{J}=\vec{N}+\vec{S}$.
The Dunham parameters $Y_{m,n}$ are equal in Hund's cases (a) and (b) for $^2\Sigma^+$ states, which can easily be seen by replacing J=N+1/2 for $e$ and J=N-1/2 for $f$ levels and the general expression N(N+1) for the rotational contribution will appear for the rotational state N in Hund's case (b). The simple energy representation was used for the uncoupled analysis adding the spin-rotation energy for a pure $^2\Sigma^+$ state with total angular momentum J or rotational angular momentum N:
	\begin{equation}
	E_\mathrm{SR} = \pm \frac{\gamma}{2}\,[N+1/2\mp 1/2]
	 = -\frac{\gamma}{2}\,[1\mp(J+1/2)],
		\label{eq:SREnergy}
	\end{equation}
where the upper sign is for \fone\ and the lower sign for \ftwo\ levels. We express the spin-rotation molecular parameter $\gamma$ with a Dunham-like expansion to include the dependence on the rovibrational motion:
	\begin{equation}
		\gamma(v,N) = \sum_{m,n} \gamma_{m,n}(v+\nicefrac{1}{2})^m[N(N+1)]^n
		\label{eq:gammaExpansion}
	\end{equation}
Fig. \ref{fig:lica_obs-cal} a) compares the spectra as evaluated using the energy expressions Eq. (\ref{eq:DunhamExpansionaSig}, \ref{eq:SREnergy} and \ref{eq:gammaExpansion}) with the experimental data. The scale J'(J'+1) is chosen to be approximately proportional to the rotational energy. The deviations are significant and systematic with respect to the experimental uncertainty asking for the development of the coupled system.

For the analysis of the coupling we use the Hamiltonian represented by the matrix in Hund's coupling case (a) with the three basis vectors $|^2\Sigma^+,v_\Sigma,J,e/f>$, $|^2\Pi_{1/2},v_\Pi,J,e/f>$, and $|^2\Pi_{3/2},v_\Pi,J,e/f>$  shown in Table \ref{tab:CouplingMatrix}.
	\begin{table*}[bth]
		\caption{\label{tab:CouplingMatrix} Matrix representation of couplings between \dSig, \dPione and \dPithree states in Hund's case (a) for a total angular momentum $J$ and the vibrational states $v_\Sigma$ and $v_\Pi$. The upper and lower signs are for $e$ and $f$ levels, respectively. See the text for the explanation of the molecular parameters.}
		\begin{center}
			\footnotesize
			\begin{tabular}{r|*3{c}} 
				& $\left|{^2\Sigma^{+}_{1/2},v_\Sigma,e/f}\right\rangle$ & $\left|{^2\Pi_{1/2},v_\Pi,e/f}\right\rangle$ & $\left|{^2\Pi_{3/2},v_\Pi,e/f}\right\rangle$  \\[10pt]
				\hline
				$\left\langle {v_\Sigma,^2\Sigma^{+}_{1/2},e/f}\right|$ & \begin{tabular}{@{}c} $E_{\mathrm{Dun}}^{\Sigma,\text{(a)}}$ \\$ -\gamma_\Sigma/2 \cdot \left [1 \mp (J+\nicefrac{1}{2}) \right ] $ \end{tabular} & \begin{tabular}{@{}c} $V_{\Sigma\Pi}\cdot p/2 \times \big[A_{\Sigma\Pi} - \gamma_{\Sigma\Pi} \, + $ \\ $2 B_{\Sigma\Pi} ( 1 \mp [J+\nicefrac{1}{2}] ) \big] $ \end{tabular}& \begin{tabular}{@{}c}$-V_{\Sigma\Pi}\cdot p \cdot B_{\Sigma\Pi}\times$\\ $\sqrt{J(J+1) -\nicefrac{3}{4}  }$ \end{tabular}\\[20pt]
				$\left\langle {v_\Pi,^2\Pi_{1/2},e/f}\right|$ & \begin{tabular}{@{}c} $V_{\Sigma\Pi}\cdot p/2 \cdot \big[A_{\Sigma\Pi} - \gamma_{\Sigma\Pi} \, + $ \\ $2 B_{\Sigma\Pi} ( 1 \mp [J+\nicefrac{1}{2}] ) \big] $ \end{tabular}  &  $E_{\mathrm{Dun}}^{\Pi,\text{(a)}}- (A_\Pi+\gamma_\Pi)/2$  & \begin{tabular}{@{}c} $(V_\Pi\cdot \gamma_\Pi/2 - B^{v}_\Pi ) \, \times $\\ $\sqrt{J(J+1) -\nicefrac{3}{4}  }$ \end{tabular} \\[20pt]
				$\left\langle {v_\Pi,^2\Pi_{3/2},e/f}\right|$ & \begin{tabular}{@{}c}$-V_{\Sigma\Pi}\cdot p \cdot B_{\Sigma\Pi}\times$\\ $\sqrt{J(J+1) -\nicefrac{3}{4}  }$ \end{tabular}  & \begin{tabular}{@{}c} $(V_\Pi\cdot \gamma_\Pi/2 - B^{v}_\Pi)  \, \times $ \\ $ \sqrt{J(J+1) -\nicefrac{3}{4}  }$ \end{tabular} &   $E_{\mathrm{Dun}}^{\Pi,\text{(a)}}+ (A_\Pi-\gamma_\Pi)/2$ 
			\end{tabular}
		\end{center}
	\end{table*}
The Hamilton operator was discussed for the case of LiSr in our previous work \cite{schwanke_laser_2017} and \cite{schwanke_2020},
 and contains the spin-orbit and rotational coupling of the two states following the development in the textbook by Lefebvre-Brion and Field  \cite{lefebvre-brion_perturbations_1986}.
	\begin{figure*}[htb]
	\includegraphics[width = 1.0\textwidth]{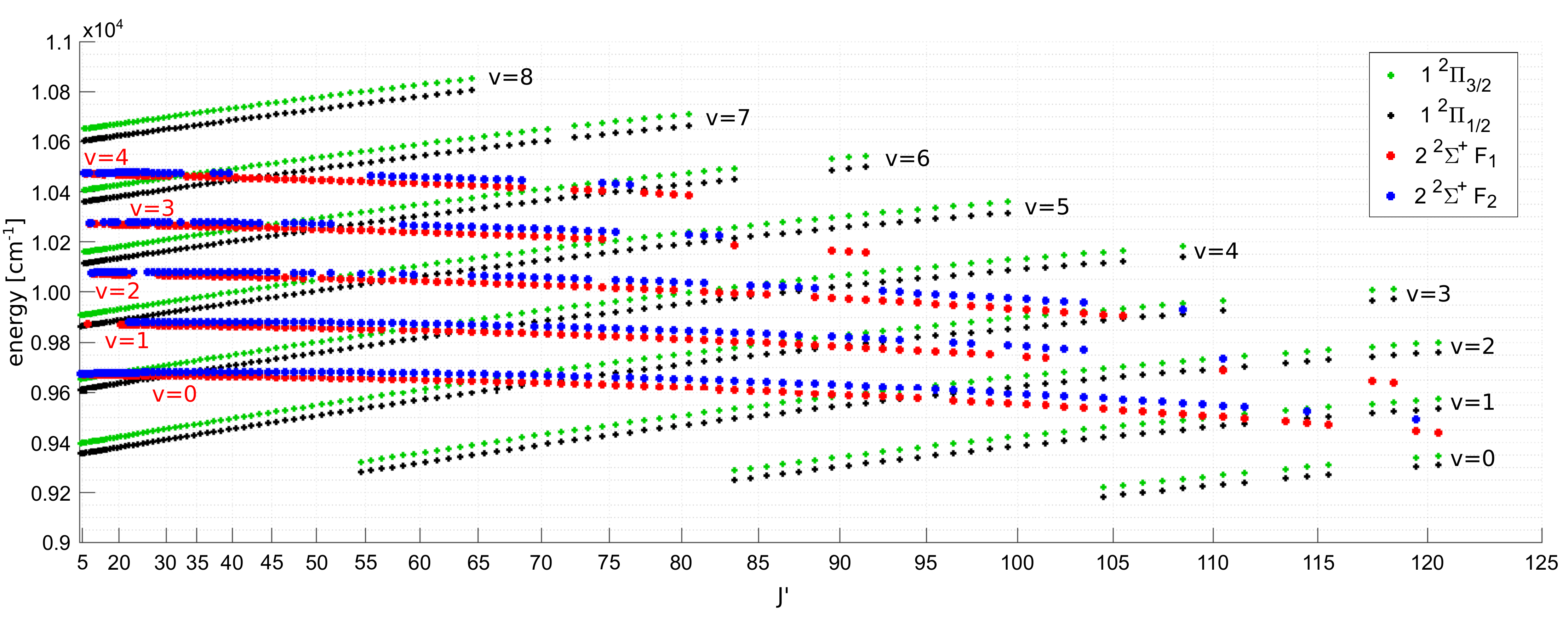}
	\caption{\label{fig:lica_Energien}
		Energy ladder of \bstate (red for $F_1$ and blue for $F_2$), $^2\Pi_{1/2}$ (black) and $^2\Pi_{3/2}$ (green) states and their crossings, indicating the expected couplings. The rotational quantum number J' is scaled with J'(J'+1) and the level energies are shifted by an averaged rotational energy [B*J'(J'+1) with B=0.23 cm$^{-1}$] to improve the visibility of the crossings.}
	\end{figure*}

For $^2\Sigma^+$ the diagonal matrix element is given by the sum of Eq. (\ref{eq:DunhamExpansionaSig}) and Eq. (\ref{eq:SREnergy}). For \dPi\ the diagonal matrix element  contains the rovibrational energy in a Dunham expansion
	\begin{equation}
		\label{eq:DunhamExpansionaPi}%
		E_\mathrm{Dun}^{\text{(a)}}(v,J) = \sum_{m,n} \mathrm{Y}^{\text{(a)}}_{m,n} \times 
		[v+\nicefrac{1}{2}]^m%
		\begin{cases}%
			[(J)(J+1)+\nicefrac{1}{4}]^n &\text{for } ^2\Pi_{1/2}\\
			[(J)(J+1)-\nicefrac{7}{4}]^n &\text{for } ^2\Pi_{3/2} ,
		\end{cases}%
	\end{equation}
the spin-orbit splitting by A$_\Pi$ and the spin-rotation contribution by $\gamma_\Pi$. This latter part is neglected not only at this place but on the whole matrix for the $\Pi$ states, because the small contribution cannot be separated from the other parts in the fitting procedure. For the final evaluation we introduce a vibrational dependence of the spin-orbit parameter in the conventional form: $A_\Pi(v)=A_\Pi^e + A_\Pi^v(v+1/2)$.
	
The rotational parameter for the non-diagonal part of the matrix in the $\Pi$ space is defined by the same Dunham parameter expansion as above
	\begin{equation}
		\label{eq:Bv}
		B_\Pi^{v} = \sum_{m} \mathrm{Y}^{\text{(a)}}_{m,1}[v+\nicefrac{1}{2}]^m
	\end{equation}
and controls the uncoupling of the spin from the molecular axis. The ab initio calculations mentioned in the introduction indicate that the spin-orbit splitting for LiCa is fairly small, thus the uncoupling starts already at low rotational levels. We treat the $\Pi$ state as an intermediate state between Hund's case (a) and (b) and the matrix of the coupled system represented in Hund's case (a) must contain the $\Pi$ state always as pairs of the $1/2$ and $3/2$ components for each vibrational level v$_\Pi$. The simplest matrix to evaluate the coupled system \bstate\ and \astate\ will be a $3\times 3$-matrix for each observed level of the state \bstate. An extension has to add vibrational levels of the $\Pi$ state in pairs of $1/2$ and $3/2$ components of equal $v_\Pi$ leading to $5\times 5$, $7\times 7$, etc. matrices.
	
The parameters for the non-diagonal matrix elements between $\Sigma$ and $\Pi$ states contain the spin-orbit (A$_{\Sigma\Pi}$)\footnote[2]{The explicit spin-rotation contribution $\gamma_{\Sigma\Pi}$ cannot be separated from the spin-orbit part.} and rotational (B$_{\Sigma\Pi}$) interaction, where $p$ stands for the expectation value of the ladder operator of the orbital angular momentum $p=<L^{\pm}>$. Assuming a weak variation of the spin-orbit interaction with internuclear separation we will model the v-dependence of this interaction by the overlap integral V$_{\Sigma\Pi}$ of the vibrational states v$_\Sigma$ and v$_\Pi$ and simplify this also for the rotational part B$_{\Sigma\Pi}$. By introducing the fit parameters
	\begin{align}
		V^\text{fit}_{\Sigma\Pi} &\coloneqq V_{\Sigma\Pi}\cdot p\cdot(A_{\Sigma\Pi}-\gamma_{\Sigma\Pi})\label{effparV}\\ 
		B^\text{fit}_{\Sigma\Pi}&\coloneqq B_{\Sigma\Pi}/(A_{\Sigma\Pi}-\gamma_{\Sigma\Pi})\label{effparB},
	\end{align}
we reduce the number of free parameters to two, where the latter one is now dimensionless and the effective overlap integral $V^\text{fit}_{\Sigma\Pi}$ has the dimension of energy.
Moreover,  $p=<L^{\pm}>$ is assumed to be the same for both states $^2\Pi_{1/2}$ and $^2\Pi_{3/2}$ and is expected to be close to $\sqrt{2}$, since these states are related to an electronic atom state with $L=1$.

The matrix in Table \ref{tab:CouplingMatrix} contains contributions of spin-rotation interaction for the state \bstate\ at two places: directly by $\gamma_\Sigma$ in the diagonal of \bstate\ summing the effect from the distant $\Pi$ states and indirectly through the non-diagonal contribution by B$_{\Sigma\Pi}$ for the close $\Pi$ states incorporated in the matrix. We will find in our analysis that the first part is dominating.

The preliminary analysis of the observations as an uncoupled system leads to the energy levels of the electronic state \bstate\ as function of the total angular momentum J', shown in Fig. \ref {fig:lica_Energien} by red and blue dots for F$_1$ and F$_2$ levels, respectively, on the scale J'(J'+1) as in Fig. \ref{fig:lica_obs-cal}. The systematic deviations in Fig. \ref{fig:lica_obs-cal} a) show the typical pattern of an avoided crossing between the directly observed levels of \bstate\ and an unknown state, here assumed as state \astate.~ An illustrative example is v'=1, in the vicinity of J'$\approx$65: The crossing of F$_1$ levels is clearly visible followed by the crossing of F$_2$ at higher J' and to lower J' still very close one sees a sharp crossing, indicating a weaker coupling than the former two. It is repeated at different J'-crossings and for different vibrational levels, getting stronger for higher v'. Its interpretation is straightforward: The stronger avoided crossing belongs to the spin-orbit coupling to \dPione\ and the weaker one to the rotational coupling between $^2\Sigma^+_{1/2}$ and \dPithree. The form of the deviations shows that the perturbing (repelling) levels come from lower energy passing from low to high J', which leads to the fact, that the rotational constant of the perturbing state will be higher than the one of state \bstate. This systematic pattern allows further conclusions: The crossing with \dPithree\ appears earlier than those for \dPione\ which means that the former energy levels are higher in energy than the latter ones. Thus the spin-orbit constant of \astate\ is positive as predicted by ab initio calculations  \cite{gopakumar_ab_2013}. The order of the strong crossings is the confirmation of the correct assignment of F$_1$ and F$_2$. For a fixed J' the level for F$_1$ of state \bstate\ has the lower rotational energy, i.e. lower N' than for F$_2$. Thus the level F$_1$ should appear first as it is the case for the chosen assignment which proves finally from experimental grounds the assignment used in ref.  \cite{ivanova_x_2011}.

The \astate\ levels needed according to the crossings in Fig. \ref{fig:lica_obs-cal} a) are constructed in such a way that their rotational ladder crosses the J' regions of the \bstate\ state where the largest deviations of the measurements appear. We start with an estimate of the vibrational spacing and rotational constant from the ab initio results of  \cite{pototschnig_electric_2016}. The needed vibrational levels of the \astate\ states are counted beginning simply by zero for the lowest state in the energy range of the observed \bstate\ levels. The ladder for \astate\ is added by black dots (\dPione) and green ones (\dPithree) to Fig. \ref {fig:lica_Energien} and the crossings show where the resonance coupling of both electronic states are expected. 
	
	\begin{table*}[bth]%
\fontsize{8pt}{8pt}\selectfont 
		\caption{Derived Dunham and spin-rotation parameters for the state \bstate\ and the state \astate\ of $^7$Li$^{40}$Ca. They give an accurate description for levels with $N\leq 120, v= 0 \text{ to } 4$ for state \bstate. The \astate\ parameters describe the vibrational structure around the observed \bstate\ levels with the assumed vibrational counting $v_\Pi = 0 \text{ to } 8$ for modeling the deperturbation.  All values are given in \si{\kayser}. For the uncertainty of the parameters see the discussion in the text.\label{tab:DunPar}}  
		
		\centering
		\setlength{\tabcolsep}{4pt}			
		\begin{tabular}{llllllr}   
			\normalsize$\dun{0}{n}$ & \normalsize{$\dun{1}{n}$} & \normalsize{$\dun{2}{n}$} & \normalsize{$\dun{3}{n}$} & \normalsize{$\dunG{0}{n}$} & \normalsize{$\dunG{1}{n}$} & \normalsize$n$ \\
			\hline
			\multicolumn{7}{l}{\normalsize \bstate \hspace{1cm} Hund's case (a)\rule{0pt}{12pt}} \\
			\hline
			$\phantom{-}9572.1011$& $\phantom{-}202.3461$    &$ -0.3383$              & $-0.020136$    &$0.0111171$     &$ -0.162\times 10^{-3}$& 0 \\
			$\phantom{-}0.2325062$& $-0.123586\times 10^{-2}$&$-0.14352\times 10^{-4}$&\phantom{-}{-}&$0.293\times 10^{-7}$&\phantom{-}{-} & 1 \\
			$-0.121683\times 10^{-5}$&$\phantom{-}0.6314\times 10^{-8}$&$\phantom{-}0.730\times 10^{-9}$&\phantom{-}{-} &\phantom{-}{-}&\phantom{-}{-}& 2 \\
			$\phantom{-} 0.691\times 10^{-11}$& \phantom{-}{-} &\phantom{-}{-} &\phantom{-}{-} &\phantom{-}{-} & \phantom{-}{-} & 3 \\
			$ -0.194\times 10^{-16}$&\phantom{-}{-} &\phantom{-}{-} &\phantom{-}{-} &\phantom{-}{-} & \phantom{-}{-} & 4 \\
			\hline
			\multicolumn{7}{l}{\normalsize \astate \hspace{1.24cm} Hund's case (a)\rule{0pt}{12pt}} \\
			\hline
			$\phantom{-}8453.734$    & $\phantom{-}269.682$     &$  -1.7461$&$\phantom{-}0.01002$&{-} & {-} & 0 \\
			$\phantom{-}0.299728$    &$-0.24997 \times 10^{-2}$ &\phantom{-}\phantom{-}$0.5115 \times 10^{-4}$&{-} &{-} & {-} & 1 \\
			$-0.4970\times 10^{-6}$  &$  -0.2199 \times 10^{-7}$&\phantom{-}{-} &{-} &{-} & {-} & 2 \\
			$-0.8709 \times 10^{-10}$&\phantom{-}{-} &\phantom{-}{-} &{-} &{-} &{-} & 3 \\
			$\phantom{-}0.2637\times 10^{-14}$&\phantom{-}{-} &\phantom{-}{-} &{-} &{-} &{-} & 4 \\
			\hline
		\end{tabular}
	\end{table*}

	\section{Fitting procedure and results}
	\label{sec:resdis}
	
For the evaluation of the coupled system we have to specify how many vibrational levels of state \astate\ are needed for each $J'$ and $v'$ of the \bstate\ state. We tried the simplest case with a single vibrational level which is the closest one to the considered level of \bstate. This leads to a ($3\times 3$)-matrix for representing \astate\ as an intermediate Hund's (a) and (b) level,  already explained above. The result was not satisfactory with an average deviation of about two times the experimental uncertainty and significant deviations around the crossings. Thus we extended the model to
 three  closest $v$ levels of the $^2\Pi_{1/2}$ and $^2\Pi_{3/2}$ states for each J' as it is shown in Fig. \ref {fig:lica_Energien}. We prefer this approach compared to only two vibrational levels. In the latter case just around each crossing a switch of the second level from above to below the crossing would appear, thus no smooth behavior is to be expected. For the case of three perturbing levels such a switch will occur for levels further away of the perturbed level; the influence on the observed levels will be reduced.
	
The parameters of the Hamiltonian in Table \ref{tab:CouplingMatrix} are determined in a nonlinear least squared fit, where for each observed \bstate\ level v', J' and parity e/f the $(7\times 7)$-matrix is diagonalized considering the three adjacent vibrational levels for each component of the \astate\ state. The fit applies the MINUIT algorithm  \cite{james_minuit_1975}.

The evaluation uses 758 different levels of state \bstate\ which are obtained from more than 3000 spectral lines, many levels are determined several times through different measured transitions. For the fit the averaged energy of each level was applied. The standard deviation of the fit is $\sigma$ = 0.012 cm$^{-1}$ with 51 fit parameters. This value is close to the average experimental uncertainty of 0.015 cm$^{-1}$. Fig. \ref{fig:lica_obs-cal} b) shows the differences between the measured values and the energies of the \bstate\ state calculated within the deperturbation. One finds that the deviations at the crossings have been significantly reduced compared to the uncoupled evaluation in Fig. \ref{fig:lica_obs-cal} a), and overall they are located within the range of experimental uncertainty which is indicated by the shaded area in the figure.
	
Table \ref{tab:DunPar} reports the derived Dunham parameters for both states \bstate\ and \astate\ and Table \ref{tab:CouplingParameters} the effective spin-orbit as well as the rotational coupling constants. 

For the effective overlap integrals $V_{\Sigma\Pi}^\text{fit}$ we discovered that it is of advantage to introduce a simple dependence on the rotational quantum number J in the neighborhood of the level crossing at $J_{cross}$ as defined in Eq. (\ref{eq:OIJ}).

	\begin{equation}
		V_{\Sigma\Pi}^\text{fit}= V_{\Sigma\Pi}^\text{const} + V_{\Sigma\Pi}^J \cdot (J-J_{cross}).
		\label{eq:OIJ}
	\end{equation}
For the values $J_{cross}$ the average of the F$_1$ and F$_2$ crossing points for each case $(v_\Sigma - v_\Pi)$ rounded to the nearest integer was selected. Introducing this extension reduces the standard deviation by a factor of 1.6 being significant in our opinion. The fitted overlap integrals are given in Table \ref{tab:OISigPi}. The $J$-dependent term $V_{\Sigma\Pi}^J$ was only applied in few cases and appears up to one order of magnitude larger than values derived from ab initio potentials. This fact might indicate that some dependence on the internuclear separation R of the spin-orbit interaction is hidden in these values. Under these circumstances, we checked if a separate fit of the rotational coupling, presently assumed to be proportional to the overlap integrals for the spin-orbit interaction (compare the definition of fit parameters in Eq. (\ref{effparB})), can help to remove the large J-dependence. The fit quality was worse and the resulting rotational coupling parameters were not physically convincing in their variation between different pairs of v$_\Sigma$ and v$_\Pi$.
	\begin{table}[bth]%
\fontsize{8pt}{8pt}\selectfont 
		\caption{Spin-orbit parameters of state \astate\ and ratio of rotational to spin-orbit coupling of the \mbox{\bstate--\astate} system. The spin-orbit coupling of  \dSig--\dPione\ is contained in the effective overlap integrals given in Table \ref{tab:OISigPi}. For the uncertainty of the parameters see the discussion in the text.  
		\label{tab:CouplingParameters}}
		\centering
		\setlength{\tabcolsep}{1pt}			
		\sisetup{
			table-number-alignment = right,
			table-figures-decimal = 5,			
			table-figures-uncertainty = 2
		}
		\begin{tabular}{@{} l S@{}}
			parameter & {value} \\
			\hline
			$A_\Pi^e$ & 37.48~ cm$^{-1}$ \\
			$A_\Pi^v$ & 1.241 ~  cm$^{-1}$\\
			$ B_{\Sigma\Pi}/(A_{\Sigma\Pi} -\gamma_{\Sigma\Pi})$ & 0.208$\times 10^{-3}$ \\
		\end{tabular}
	\end{table}
	
	\begin{table*}[bth]%
\fontsize{8pt}{8pt}\selectfont 
			
			\caption{\label{tab:OISigPi} Overlap integrals $V_{\Sigma\Pi}^\text{fit}$ between the \bstate\ and \astate\ states from the
			deperturbation. The overlap integrals are in unit \si{\kayser}, because they contain the spin-orbit interaction.
				The values $V_{\Sigma\Pi}^\text{const}$ and $V_{\Sigma\Pi}^J$ belong to the power expansion with  $J_{cross}$ according to Eq. (\ref{eq:OIJ}). See the text for the estimate of the uncertainties.
				}     %

		\begin{minipage}{0.9\textwidth}  
			\centering
			 \setlength{\tabcolsep}{2pt}			
			\sisetup{
				table-number-alignment = right,
				table-figures-decimal = 4,			
			}
			\begin{tabular}{@{}
					c|
					S[table-figures-exponent = 0,table-figures-decimal = 2] 
					S[table-figures-exponent = 0,table-figures-decimal = 0]
					c|
					S[table-figures-exponent = 0,table-figures-decimal = 2]
					S[table-figures-exponent = 0,table-figures-decimal = 2]
					c|
					S[table-figures-exponent = 0,table-figures-decimal = 2]
					S[table-figures-exponent = 0,table-figures-decimal = 2]
					c|
					S[table-figures-exponent = 0,table-figures-decimal = 2]
					S[table-figures-exponent = 0,table-figures-decimal = 2]
					c|
					S[table-figures-exponent = 0,table-figures-decimal = 2]
					S[table-figures-exponent = 0,table-figures-decimal = 2]
					c
					@{}}
				$v_\Pi$& \multicolumn{3}{c|}{$v_\Sigma=0$}  & \multicolumn{3}{c|}{$v_\Sigma=1$}   & \multicolumn{3}{c|}{$v_\Sigma=2$}   & \multicolumn{3}{c|}{$v_\Sigma=3$}  & \multicolumn{3}{c}{$v_\Sigma=4$}   \\
				&$V_{\Sigma\Pi}^\text{const}$       & $V_{\Sigma\Pi}^J$  & $J_{cross}$ & $V_{\Sigma\Pi}^\text{const}$ & $V_{\Sigma\Pi}^J$ &$J_{cross}$ & $V_{\Sigma\Pi}^\text{const}$   & $V_{\Sigma\Pi}^J$ &$J_{cross}$& $V_{\Sigma\Pi}^\text{const}$     & $V_{\Sigma\Pi}^J$   &$J_{cross}$& $V_{\Sigma\Pi}^\text{const}$    & $V_{\Sigma\Pi}^J$ &$J_{cross}$ \\
				\hline
				0 & 3.3\textsuperscript{a} &{-}&$>$130  &{-}&{-}&{-}  &{-}&{-}&{-}  &{-}&{-}&{-}  &{-}&{-}&{-} \\  
				1 &2.988&{-}&115 &11$^a$&{-}&$>$130 &{-}&{-}&{-} &{-}&{-}&{-} &{-}&{-}&{-} \\ 
				2 &1.592&{-}   &95   &8.158   &0.081&112 &5.2$^a$&{-}  &$>130$&{-}&{-}&{-}    &{-}&{-}&{-} \\
				3 &0.816&{-}   &71   &3.959   &0.074&91  &12.47  &0.25 &110   &{-}&{-}&{-}    &{-}&{-}&{-} \\
				4 &0.29$^a$&{-}&33   &2.008   &{-}  &65  &7.15   &0.12 &86    &18.76&0.65&104 &{-}&{-}&{-} \\
				5 &0.1$^a$&{-} &$<$10&0.87$^a$&{-}  &14  &4.03   &{-}  &58    &10.28&0.12&75  &2.4$^{a}$&{-}  &93 \\
				6 &{-}&{-}&{-}       &0.33$^a$&{-}  &{b} &1.8$^a$&{-}  &14    &6.59&{-}  &51  &15.00    &0.031&75 \\
				7 &{-}&{-}&{-}       &{-}     &{-}  &{-} &0.74$^{a}$&{-}&{b}  &3.0$^a$&{-}&14 &9.62     &0.051&43 \\
				8 &{-}&{-}&{-}       &{-}     &{-}  &{-} &{-}&{-}&{-}         &1.0$^a$&{-}&{b}&5.2$^a$  &{-}  &{b}  \\
			\end{tabular}
				\footnotetext[1]{Parameters with label a were kept fixed during final fit.}
			\footnotetext[2]{The label b indicates that the crossing is beyond our data range or not existing (compare Figure \ref{fig:lica_Energien}).}
		\end{minipage}
	\end{table*}
	
The number of introduced fit parameters is fairly high and from the many trials during the evaluation we learned, that the correlation between them is high. Thus we give no uncertainties for the individual parameters, which could lead to overinterpretation. The parameters will allow to recalculate the studied levels to an accuracy of 0.012 cm$^{-1}$, the standard deviation of the fit, and an extrapolation to levels not too far from the quantum number regime studied here. Any extrapolation to perturbed ranges not incorporated here will be dangerous. This is the notable limitation of the local deperturbation within a specific range of vibrational levels.
	
For the Dunham description of the molecular levels we used centrifugal distortion parameters up to Y$_{04}$, this is dictated by the desire to obtain a good fit but produces a strong J'-dependence, and a significant correlation between the corresponding parameters. A different approach can be helpful to change the correlation or hopefully to reduce it, namely the use of a molecular potential instead of molecular Dunham parameters. For the state \bstate\ we studied a continuous vibrational ladder from v=0 to 4, which is advantageous for deriving a molecular potential in the corresponding energy range. But for state \astate\ we have only a set of vibrational levels, missing the vibrational assignment and therefore information for building up the potential from the bottom, which is expected from ab initio calculations to be much lower than the observed region. Thus we studied the potential approach for the state \bstate.

For modeling the potential of \bstate\ we use the conventional analytical description  \cite{ivanova_x_2011}  in a finite range of internuclear separation R$_{in} < R < R_{out}$: 
\begin{equation}
\label{eq:uanal}
U(R)=\sum_{i=0}^{n}a_i\,\xi(R)^i
\end{equation}
with the expansion parameter
\begin{equation}
\label{eq:xv}
\xi(R)=\frac{R - R_m}{R + b\,R_m},
\end{equation}
\noindent where the coefficients $a_i$ are fitted
parameters and $b$ and $R_m$ are fixed. $R_m$ is normally close to the value of the
equilibrium separation $R_e$ (for more detail see  \cite{Samuelis:00}). The repulsive branch of the potential is extrapolated for $R <R_{\rm in}$ with:
\begin{equation}
\label{eq:rep}
  U(R)= A + B/R^{N_s}
\end{equation}
\noindent adjusting the parameters $A$, $B$ with $N_s=6$ to get a smooth
transition at $R_{\rm in}$, thus these are no free parameters of the fit later on.

For large internuclear separations ($R > R_{\rm out}$)
we adopted the standard long range form with dispersion coefficients C$_n$:
\begin{equation}
U_{\mathrm{LR}}=U_{\infty}-C_6/R^6-C_8/R^8-C_{10}/R^{10}
\label{lr}
 \end{equation}
the values of the C$_n$ and the asymptotic energy U$_{\infty}$ are of no importance in our case, because we evaluate only low vibrational levels and thus never approach the long range region. The $C_8$ and $C_{10}$ are used for the smooth connection at R$_{out}$. For the others we introduce values to be consistent with ab initio results and the asymptotic atom pair correlation.

The rovibrational level for the desired quantum state v', N'=J' $+/- 1/2$ and $F_{2/1}$ is calculated by solving the one-dimensional Schr\"odinger equation for such a potential. The eigen energy replaces the one of the Dunham description in the matrix in Table \ref{tab:CouplingMatrix}. Using in total 9 potential parameters instead of 11 Dunham parameters we obtain a good fit with a standard deviation of $\sigma$=0.012 cm$^{-1}$ which is equal to the one with the pure Dunham approach. The differences of the two rovibrational ladders as function of N' undulate around zero with an amplitude of about 0.01 cm$^{-1}$ but increase to larger values for high N', which is probably related to the high order centrifugal distortion function constructed by the Dunham approach. We should mention that the other molecular parameters are also slightly changed by the simultaneous fit of all parameters directing to the correlation between all parameters which is different in both approaches. The parameters from the potential approach are given in the appendix. 

In order to determine the physically interesting spin-orbit coupling $A_{\Sigma\Pi}$ from fitted effective overlap integrals $V^\text{fit}_{\Sigma\Pi}$, we compare their relative variation with that of the overlap integrals $V_{\Sigma\Pi} $ calculated from the ab initio potentials  \cite{pototschnig_vibronic_2017}. 
First, the comparison of the energetic position of the vibrational state of the \astate\ state, denoted by $v^{fit}=0$, with the ab initio calculations  \cite{pototschnig_vibronic_2017}  suggests a shift of 11, i.e. $v=v^{fit}+11$. Additionally, the distribution of the fitted overlap integrals shows the highest similarity with variation of those derived from the ab initio potentials  \cite{pototschnig_vibronic_2017}  for the same vibrational shift. This procedure provides the appropriate $v$ assignment of the indirectly observed $^2\Pi$ levels. The ratio of the two series of overlap integrals (the first one fitted with observations and the second one from ab initio potentials) yields as averaged value over all vibrational pairs where crossings were observed
\begin{equation}
	(\frac{V_{\Sigma\Pi}^\text{fit}}{V_{\Sigma\Pi}^\text{calc}})_{averaged}= 82.5 \si{\kayser}
	\label{eq:Vratio}
\end{equation}
 with a statistical spread of 5\%. Using Eq. (\ref{effparV}) and assuming that $\gamma_{\Sigma \Pi}$ is much smaller than $A_{\Sigma \Pi}$ and $p=\sqrt 2$, the constant $A_{\Sigma \Pi}$  can be estimated 

\begin{equation}
	A_{\Sigma\Pi} \approx  ( \frac{V_{\Sigma\Pi}^\text{fit}}{V_{\Sigma\Pi}^\text{calc}})_{averaged} \frac{1}{\sqrt{2}} = 58.3 \si{\kayser}.
	\label{eq:Avalue}
\end{equation}
The spin-orbit constants $A_{\Sigma\Pi}$ and $A_{\Pi}$ (Table \refeq{tab:CouplingParameters}) are of similar magnitude giving support to this estimate and the vibrational assignment. 
Similarly, Eq. (\ref{effparB}) allows to estimate
\begin{equation}
	B_{\Sigma\Pi} = 0.012 \si{\kayser}
	\label{eq:Bvalue}
\end{equation}	
for the rotational coupling between $\Sigma$ and $\Pi$. This effective rotational constant is much smaller than the rotational constants of states \astate\ and \bstate\ leading to a small contribution to the spin-rotation splitting of state \bstate\ compared to the direct part by the $\gamma$-parameter in the matrix.

The table of the effective overlap integrals Tab. \ref{tab:OISigPi} shows finite values at places where no crossing of the perturbing levels is observed. They were fixed during the fit. A first estimate of these values was obtained from the calculated overlap integrals multiplied by the ratio derived in Eq. (\ref{eq:Vratio}) above. They were adjusted slightly to lower values by hand because the coupling to energetically far lying states was too strong to fit the observations. A free fit led to unphysical values.

	\section{Discussion and Outlook}
	\label{outlook}
The energy levels of the \bstate\ state of LiCa have been measured up to  v'=4  and the Dunham model and the molecular potential of this state has been significantly extended compared to ref.  \cite{stein_spectroscopic_2013}. Due to the discovered local perturbations caused by the interaction with the \astate\ state, a Dunham  description of that state was derived, as well as the coupling constants with the state \bstate. The final deperturbation model was applied to simulate the thermal emission spectrum as shown in Fig.\ref{fig:lica_sim} b). With the precise knowledge of the ground state and the 758 different excited levels evaluated in this study the range of spectral lines covering more than 10000 different lines was extended with ground state levels from v"=0 to 6. Bands with higher v" can be predicted from the knowledge of the ground state potential. Not all of the calculated lines are detectable within the sensitivity limits of our experiment, but many places were checked and clearly identified in the observed spectrum. For example, almost all weak features seen in Fig. \ref{fig:lica_lif} c) are contained in the simulation. The consistency is very satisfying and during this examination lines resulting to about 10\% new excited levels were assigned. This assignment is justified because the identified lines are well separated despite the otherwise dense overlapping spectrum. This in total proves the reliability of the present analysis. Few lines belonging to $^6$Li$^{40}$Ca were detected during our intensive study, but the information is too little to investigate Born-Oppenheimer corrections between the two isotopologues. We also were looking for extra lines resulting from the perturbation by \astate, but due to the dense spectrum no unambiguous assignment of such lines was successful.

\begin{table*}[htb]
		\caption{Comparison of measured spectroscopic constants of $^7$Li$^{40}$Ca with results of the Graz group \cite{pototschnig_vibronic_2017,pototschnig_private_2017} (MRCI).	For the \astate\ state, we calculated the parameters for $v_\Pi =11$ from the ab initio potential and compare these with $v=0$ of our local description. All values are given in cm$^{-1}$, except $R$ in \AA. \label{tab:constComp}}	
			\footnotesize
			\noindent
			\begin{minipage}{1.0\textwidth}
				\begin{tabularx}{0.9\textwidth}{@{}XYYYYYr@{}}
					\hline
					\noalign{\vskip 5pt}   
					&{$R_e$} &{$\omega_e = \dun{1}{0}$}&{$B_e = \dun{0}{1}$} &{$T_e $} &  & \\
					\hline
					\noalign{\vskip 5pt}   
					\small{\bstate}& 3.518$\phantom{00}$ & 204.7\phantom{000} & - & 9570$\phantom{.00}$ & & ab initio\\
					& 3.4854\textsuperscript{a} & 202.368 \phantom{0}&  0.23252$\phantom{^\text{b}}$ & 9572.09& & this work\\
					\hline
					\hline
					\noalign{\vskip 5pt}
					& {$R$} &{$E(11)-E(10)\approx \dun{1}{0}$}&{$B_{11}\approx \dun{0}{1}$} &{$E(11) \approx E(v^{Fit}=0)$} & {$A_\Pi$} & \\
					\hline
					\noalign{\vskip 5 pt}
					\small{\astate}& 2.990\textsuperscript{b} &  262.4\textsuperscript{b}$\phantom{00}$ & 0.293\textsuperscript{b}$\phantom{00}$  & 8641\textsuperscript{b}  & 36.65\textsuperscript{c} & ab initio\\
					& 2.906\textsuperscript{a} & 269.682$\phantom{^\text{b}}$ & 0.2997$\phantom{0^\text{b}}$ &8588.1 & 37.48$\phantom{^\text{a}}$ & this work\\   
				\end{tabularx}
			\footnotetext[1]{Calculated from \dun{0}{1}}
			\footnotetext[2]{Calculated with MRCI potential \cite{pototschnig_vibronic_2017,pototschnig_private_2017} for $v=11$}
			\footnotetext[3]{Taken from Ref. \cite{gopakumar_ab_2013} (SO-MS-CASPT2). The other molecular constants from Ref. \cite{gopakumar_ab_2013} agree within ca. \SI{7}{\percent} with those of Ref. \cite{pototschnig_vibronic_2017}.}
			
		\end{minipage}%
	\end{table*}

Expanding the description to higher $v$ levels of the \bstate\ state would be desirable. The corresponding lines and band heads appear in the thermal spectrum above 10000 \si{\kayser} (see Fig. \ref{fig:lica_sim}), but they are much weaker than the lines already examined, and this will lead to  difficulties for a successful laser induced fluorescence experiment. 	

In Table \ref{tab:constComp} we compare the experimentally determined molecular parameters of the states \bstate\ and \astate\ with the results of the ab initio work  \cite{pototschnig_vibronic_2017}. For the state \bstate\ the energetic positions given as $T_e$ are almost equal but the potentials are slightly shifted on the internuclear distant axis, compare the values of $R_e$ for the position of the minimum of the potential. The actual forms of the potentials are very similar, which is consistent with the close agreement in the vibrational constant $\omega_e$. 

Since the \astate\ state was not directly measured spectroscopically and only a part of the vibrational ladder could be investigated, we calculate with the MRCI potentials  \cite{pototschnig_vibronic_2017}  the vibrational spacing and the rotational constant for v=11 as assigned from the overlap integrals to compare these values with those from our analysis, arbitrarily assigned to v=0. The difference between the fitted energy $E(v^{Fit}=0)$ and the theoretical one $E(v=11)$ is only 53 \si{\kayser}. This is not as good as in the case of \bstate\ but it is significantly smaller than the vibrational spacing, which supports the vibrational assignment. Rotational constants and vibrational spacings differ by less than \SI{3}{\percent} between experiment and theory. This is a good starting point for using the ab initio potentials in a coupled channel calculation with radial functions for spin-orbit and spin-rotation interactions. This is a long term goal of investigating the electronic system of LiCa and needs extensive experimental work to find the state \astate\ by its spectrum and not only indirectly by perturbations. But ab initio work \cite{pototschnig_electric_2016,gopakumar_ab_2013} predicts that the transition dipole moment for \astate\ - \Xstate\ transitions is fairly low which asks for significant improvement in the detection sensitivity.

The derived spin-orbit constant as shown in Tab. \ref{tab:CouplingParameters} or \ref{tab:constComp} is very close to the ab initio result in Ref. \cite{gopakumar_ab_2013}. We should note that the spin-orbit constant $A_\Pi$ correlates strongly with the other parameters like \dun{1}{0} from the $\Pi$ state. Thus we found fits with similar standard deviations as reported here covering a range $A_\Pi=39\pm3$ \si{\kayser} of the spin-orbit parameter. Like this large uncertainty of the spin-orbit interaction the derived Dunham parameters of state \astate\ are only effective parameters representing the rovibrational level structure shown in Fig. \ref{fig:lica_Energien}. The solid information on the state \astate\ obtained in this study are the positions of the level crossings between states \bstate\ and \astate, i.e. the energetic positions of the resonance coupling.

The stronger J-dependence of the overlap integrals derived from the observations compared to ab initio calculations might be caused by correlations between these parameters and the Dunham coefficients of the two involved states. However, deperturbation with the potential description of the \bstate\ state yielded similar J-dependencies of the overlap integrals. Another reason could be the R-dependence of the spin orbit interaction, which might manifest itself in the J-dependence of the overlap integrals through which any R-dependence was neglected by definition.

The spin-rotation constant $\gamma$ of the state \bstate\ was clearly needed for a good fit and the spin-rotation splitting influenced by the spin-orbit interaction with the \astate\ state was not sufficient. Obviously, the other more distant molecular states with $\Pi$ character make a major contribution to the spin-rotation energy of state \bstate.

Analogously to the molecules KCa \cite{gerschmann_laser_2017}, LiSr \cite{schwanke_laser_2017,schwanke_2020}  and LiCa investigated in this study, other alkali-metal alkaline-earth-metal molecules can be investigated in the same way. For example, the RbSr molecule currently used to generate ultracold molecules  \cite{pasquiou2013,devolder2021}  can be produced analogously in a heat pipe and examined with high-resolution spectroscopy using LIF experiments. The thermal emission spectrum of this molecule was already detected by our group  and recently, its ground state was spectroscopically studied by Ciamei et al \cite{ciamei_rbsr_2018}. As with LiSr and LiCa, the \astate\ state of RbSr could be analyzed via deperturbation, since this state is difficult to observe directly due to a low transition dipole moment between the \astate\ and ground state of the alkali-metal alkaline-earth-metal molecules  \cite{pototschnig_ab_2014}. Feshbach resonances were also observed in collisions of ultracold Rb and Sr \cite{barbe_observation_2018}.	
	
	
\textit{Acknowledgments:} We submit this article in honor of Wim Ubachs on the occasion of his 65$^{th}$ birthday. We thank Horst Kn\"ockel and Asen Pashov for many helpful discussions and advice during the course of this research.

	
	\bibliographystyle{tfo}
	\bibliography{Bibliography_LiCa_2022}

\begin{thebibliography}{24}
\providecommand{\url}[1]{\texttt{#1}}
\providecommand{\urlprefix}{URL }

\bibitem{kotochigova_ab_2011}
S. Kotochigova, A. Petrov, M. Linnik, J. Kłos and P.S. Julienne,  The Journal
  of Chemical Physics  \textbf{135} (16), 164108 (2011).

\bibitem{Zhu:20}
Z.X. Ye, L.Y. Xie, Z. Guo, X.B. Ma, G.R. Wang, L. You and M.K. Tey,  Phys. Rev.
  A  \textbf{102}, 033307 (2020).

\bibitem{barbe_observation_2018}
V. Barbé, A. Ciamei, B. Pasquiou, L. Reichsöllner, F. Schreck, P.S.
  Żuchowski and J.M. Hutson,  Nature Physics  \textbf{14} (9), 881--884
  (2018).

\bibitem{Hara:11}
H. Hara, Y. Takasu, Y. Yamaoka, J.M. Doyle and Y. Takahashi,  Phys. Rev. Lett.
  \textbf{106}, 205304 (2011).

\bibitem{Green:20}
A. Green, H. Li, J.H.S. Toh, X. Tang, K.C. McCormick, M. Li, E. Tiesinga, S.
  Kotochigova and S. Gupta,  {Phys. Rev. X}  \textbf{10}, 031037 (2020).

\bibitem{pototschnig_electric_2016}
J.V. Pototschnig, A.W. Hauser and W.E. Ernst,  Physical Chemistry Chemical
  Physics  \textbf{18} (8), 5964--5973 (2016).

\bibitem{pototschnig_vibronic_2017}
J.V. Pototschnig, R. Meyer, A.W. Hauser and W.E. Ernst,  Physical Review A
  \textbf{95} (2), 022501 (2017).

\bibitem{gopakumar_ab_2013}
G. Gopakumar, M. Abe, M. Hada and M. Kajita,  The Journal of Chemical Physics
  \textbf{138} (19), 194307 (2013).

\bibitem{gopakumar_dipole_2014}
G. Gopakumar, M. Abe, M. Hada and M. Kajita,  The Journal of Chemical Physics
  \textbf{140} (22), 224303 (2014).

\bibitem{ivanova_x_2011}
M. Ivanova, A. Stein, A. Pashov, A.V. Stolyarov, H. Knöckel and E. Tiemann,
  The Journal of Chemical Physics  \textbf{135} (17), 174303 (2011).

\bibitem{stein_spectroscopic_2013}
A. Stein, M. Ivanova, A. Pashov, H. Knöckel and E. Tiemann,  The Journal of
  Chemical Physics  \textbf{138} (11), 114306 (2013).

\bibitem{schwanke_laser_2017}
E. Schwanke, H. Knöckel, A. Stein, A. Pashov, S. Ospelkaus and E. Tiemann,
  Journal of Physics B: Atomic, Molecular and Optical Physics  \textbf{50}
  (23), 235103 (2017).

\bibitem{schwanke_2020}
E. Schwanke, J. Gerschmann, H. Knöckel, S. Ospelkaus and E. Tiemann,  Journal
  of Physics B: Atomic, Molecular and Optical Physics  \textbf{53} (6), 065102
  (2020).

\bibitem{gerschmann_laser_2017}
J. Gerschmann, E. Schwanke, A. Pashov, H. Knöckel, S. Ospelkaus and E.
  Tiemann,  Physical Review A  \textbf{96} (3), 032505 (2017).

\bibitem{sup}
supplementary material: LiCa-spectrum.dat and LiCa-lines.dat .

\bibitem{lefebvre-brion_perturbations_1986}
H. Lefebvre-Brion and R.W. Field, \emph{Perturbations in the spectra of
  diatomic molecules}   (Academic Press, Orlando, 1986).

\bibitem{james_minuit_1975}
F. James and M. Roos,  Computer Physics Communications  \textbf{10} (6),
  343--367 (1975).

\bibitem{Samuelis:00}
C. Samuelis, E. Tiesinga, T. Laue, M. Elbs, H. {Kn\"{o}ckel} and E. Tiemann,
  Phys. Rev. A  \textbf{63}, 012710 (2000).

\bibitem{pototschnig_private_2017}
J.V. Pototschnig, R. Meyer, A.W. Hauser and W.E. Ernst,  personal communication
    (2017).

\bibitem{pasquiou2013}
B. Pasquiou, A. Bayerle, S.M. Tzanova, S. Stellmer, J. Szczepkowski, M.
  Parigger, R. Grimm and F. Schreck,  Phys. Rev. A  \textbf{88}, 023601 (2013).

\bibitem{devolder2021}
A. Devolder, M. Desouter-Lecomte, O. Atabek, E. Luc-Koenig and O. Dulieu,
  Phys. Rev. A  \textbf{103}, 033301 (2021).

\bibitem{ciamei_rbsr_2018}
A. Ciamei, J. Szczepkowski, A. Bayerle, V. Barbé, L. Reichsöllner, S.M.
  Tzanova, C.C. Chen, B. Pasquiou, A. Grochola, P. Kowalczyk, W. Jastrzebski
  and F. Schreck,  Physical Chemistry Chemical Physics  \textbf{20} (41),
  26221--26240 (2018).

\bibitem{pototschnig_ab_2014}
J.V. Pototschnig, G. Krois, F. Lackner and W.E. Ernst,  The Journal of Chemical
  Physics  \textbf{141} (23), 234309 (2014).

\bibitem{Sansonetti:11}
C.J. Sansonetti, C. Simien, J. Gillaspy, J.N. Tan, S.M. Brewer, R.C. Brown, S.
  Wu and J. Porto,  Phys. Rev. Lett.  \textbf{107}, 023001 (2011).

\end{thebibliography}

\section{Appendix}

In Table \ref{pot_b}, we present the potential parameters and include the spin-rotation parameters derived for the state \bstate.
The Dunham parameters of the state \astate\ obtained from a simultaneous fit of the potential and the effective overlap integrals can be found in Tables \ref{DunPar_pot}, \ref{tab:CP_pot} and \ref{tab:OI_pot}. The definition of these parameters is equal to the case of former Dunham approach. The asymptotic energy $U_\infty$ was calculated using the dissociation energy of the ground state Ref. \cite{ivanova_x_2011}  and adding the atomic energy difference Li($^2P$)-Li($^2S$)  \cite{Sansonetti:11}, because this molecular state will adiabatically correlate to the atom-pair asymptote Li($^2P$)+Ca($^1S$). We do not need to consider the atomic spin-orbit splitting here, because it is much smaller than the accuracy of the ground state dissociation energy.
\begin{table}[h]
\fontsize{8pt}{8pt}\selectfont 
\caption{Parameters of the
analytic representation of the state \bstate\ potential and the spin-rotation parameters for $^7$Li$^{40}$Ca. The energy reference is the minimum of the ground state potential. Parameters with an asterisk $^\ast$ ensure smooth extrapolation of the potential at $R_{in}$.\label{pot_b}} 
\vspace{3mm}
\begin{tabular*}{0.8\linewidth}{@{\extracolsep{\fill}}ccc}
\hline
\multicolumn{3}{c}{$R < R_\mathrm{in}= 2.90$ \AA} \\
\hline
$A^\ast$ & $0.722247892\times 10^{4}$ &\wn \\
$B^\ast$ & $0.247536030\times 10^{7}$ &\wn \AA $^{6}$ \\
\hline
\multicolumn{3}{c}{$R_\mathrm{in} \leq R \leq R_\mathrm{out}=  4.50$ \AA} \\
\hline
$b$ & $-0.40$ \\
$R_\mathrm{m}$ & $3.48535$& \AA \\
 $a_{0}$ & $9571.97780$&\wn\\
 $a_{1}$ & $ 3.26745027082784434$& \wn\\
 $a_{2}$ & $ 0.158554707744089319\times 10^{ 5}$& \wn\\
 $a_{3}$ & $ 0.145340686470776764\times 10^{ 5}$& \wn\\
 $a_{4}$ & $ 0.139210958410380827\times 10^{ 5}$& \wn\\
 $a_{5}$ & $ 0.966642945168972074\times 10^{ 4}$& \wn\\
 $a_{6}$ & $-0.495415350882863204\times 10^{ 5}$& \wn\\
 $a_{7}$ & $-0.726251102202838956\times 10^{ 5}$& \wn\\
 $a_{8}$ & $ 0.216117056448716437\times 10^{ 6}$& \wn\\
\hline
\multicolumn{3}{c}{$R_\mathrm{out} < R$}\\
\hline
${U_\infty}$ & 17509.20 \wn \\
${C_6}$    &  0.5995116$\times 10^{8}$ &\wn\AA$^6$ \\
${C_{8}^\ast}$  &  0.9312731$\times 10^{9}$ &\wn\AA$^8$ \\
${C_{10}^\ast}$ & -0.2436558$\times 10^{11}$& \wn\AA$^{10}$ \\
\hline
\multicolumn{3}{c}{spin-rotation in Dunham notation}\\
\hline
$\gamma_{00}$ & 0.110804$\times 10^{-1}$  & \wn \\
$\gamma_{10}$ &-0.149$\times 10^{-3}$ &  \wn\\
$\gamma_{01}$ & 0.379$\times 10^{-7}$ & \wn\\
\hline
\hline
\multicolumn{3}{c}{Derived constant:} \\
\hline
\multicolumn{3}{l}{equilibrium separation:\hspace{0.5cm} $R_e$= 3.48513(10) \AA} \\
\hline
\end{tabular*}
\end{table}
\begin{table}[bth]%
\fontsize{8pt}{8pt}\selectfont 
		\caption{Derived Dunham coefficients for the state \astate\ of $^7$Li$^{40}$Ca obtained by the potential approach. All values are given in \si{\kayser}. \label{DunPar_pot}}  
		\centering
		\setlength{\tabcolsep}{3pt}			
		\begin{tabular}{llllr}   
			
			\normalsize$\dun{0}{n}$ & \normalsize{$\dun{1}{n}$} & \normalsize{$\dun{2}{n}$} & \normalsize{$\dun{3}{n}$} & \normalsize$n$ \\
			
			\hline
			\multicolumn{5}{l}{\normalsize \astate \hspace{1.24cm} Hund's case (a)\rule{0pt}{12pt}} \\
			\hline
			$\phantom{-}8457.696$    & $\phantom{-}269.682$     &$-1.8827$&$\phantom{-}0.01962$& 0 \\
			$\phantom{-}0.299411$    &$-0.26332 \times 10^{-2}$ &\phantom{-}\phantom{-}$0.6436 \times 10^{-4}$&{-} & 1 \\
			$-0.4970\times 10^{-6}$  &$  -0.8984 \times 10^{-8}$&\phantom{-}{-} &{-}  & 2 \\
			$-0.9135 \times 10^{-10}$&\phantom{-}{-} &\phantom{-}{-} &{-}  & 3 \\
			$\phantom{-}0.2940\times 10^{-14}$&\phantom{-}{-} &\phantom{-}{-} &{-}  & 4 \\
			\hline
		\end{tabular}
	\end{table}
\begin{table}[bth]%
\fontsize{8pt}{8pt}\selectfont 
		\caption{Spin-orbit parameters of state \astate\ and ratio of rotational to spin-orbit coupling of the \mbox{\bstate--\astate} system from the potential approach.
		\label{tab:CP_pot}} 
		\centering
		\setlength{\tabcolsep}{1pt}			
		\sisetup{
			table-number-alignment = right,
			table-figures-decimal = 5,			
			table-figures-uncertainty = 2
		}
		\begin{tabular}{@{} l S@{}}
			parameter & {value} \\
			\hline
			$A_\Pi$ & 38.44~ cm$^{-1}$ \\
			$A_\Pi^v$ & 1.004 ~  cm$^{-1}$\\
			$ B_{\Sigma\Pi}/(A_{\Sigma\Pi} -\gamma_{\Sigma\Pi})$ & 0.207$\times 10^{-3}$ \\
			\hline
		\end{tabular}
	\end{table}
\begin{table*}[bth]%
\fontsize{8pt}{8pt}\selectfont 
			\caption{\label{tab:OI_pot} Overlap integrals $V_{\Sigma\Pi}^\text{fit}$  between the \bstate\ and \astate\ states from the deperturbation with the potential approach. Units of \si{\kayser}.
				}     

		\begin{minipage}{1.0\textwidth}  
			\centering
			 \setlength{\tabcolsep}{2pt}			
			\sisetup{
				table-number-alignment = right,
				table-figures-decimal = 4,			
			}
			\begin{tabular}{@{}
					c|
					S[table-figures-exponent = 0,table-figures-decimal = 2] 
					S[table-figures-exponent = 0,table-figures-decimal = 0]
					c|
					S[table-figures-exponent = 0,table-figures-decimal = 2]
					S[table-figures-exponent = 0,table-figures-decimal = 2]
					c|
					S[table-figures-exponent = 0,table-figures-decimal = 2]
					S[table-figures-exponent = 0,table-figures-decimal = 2]
					c|
					S[table-figures-exponent = 0,table-figures-decimal = 2]
					S[table-figures-exponent = 0,table-figures-decimal = 2]
					c|
					S[table-figures-exponent = 0,table-figures-decimal = 2]
					S[table-figures-exponent = 0,table-figures-decimal = 2]
					c
					@{}}
				$v_\Pi$& \multicolumn{3}{c|}{$v_\Sigma=0$}  & \multicolumn{3}{c|}{$v_\Sigma=1$}   & \multicolumn{3}{c|}{$v_\Sigma=2$}   & \multicolumn{3}{c|}{$v_\Sigma=3$}  & \multicolumn{3}{c}{$v_\Sigma=4$}   \\
				&$V_{\Sigma\Pi}^\text{const}$       & $V_{\Sigma\Pi}^J$  & $J_{cross}$ & $V_{\Sigma\Pi}^\text{const}$ & $V_{\Sigma\Pi}^J$ &$J_{cross}$ & $V_{\Sigma\Pi}^\text{const}$   & $V_{\Sigma\Pi}^J$ &$J_{cross}$& $V_{\Sigma\Pi}^\text{const}$     & $V_{\Sigma\Pi}^J$   &$J_{cross}$& $V_{\Sigma\Pi}^\text{const}$    & $V_{\Sigma\Pi}^J$ &$J_{cross}$ \\
				 & \multicolumn{3}{c|}{ }&\multicolumn{3}{c|} { }&\multicolumn{3}{c|}{ } &\multicolumn{3}{c|}{ } &\multicolumn{3}{c}{ }  \\
				\hline
				0 & 3.3\textsuperscript{a} &{-}&$>$130  &{-}&{-}&{-}  &{-}&{-}&{-}  &{-}&{-}&{-}  &{-}&{-}&{-} \\  
				1 &2.896&{-}&115 &11$^a$&{-}&$>$130 &{-}&{-}&{-} &{-}&{-}&{-} &{-}&{-}&{-} \\ 
				2 &1.709&{-}   &95   &8.003   &0.299&112 &5.2$^a$&{-}  &$>130$&{-}&{-}&{-}    &{-}&{-}&{-} \\
				3 &0.651&{-}   &71   &3.963   &0.056&91  &13.31  &0.32 &110   &{-}&{-}&{-}    &{-}&{-}&{-} \\
				4 &0.29$^a$&{-}&33   &2.043   &{-}  &65  &7.21   &0.076 &86    &18.37&0.49&104 &{-}&{-}&{-} \\
				5 &0.1$^a$&{-} &$<$10&0.87$^a$&{-}  &14  &4.07   &{-}  &58    &10.16&0.14&75  &2.4$^{a}$&{-}  &93 \\
				6 &{-}&{-}&{-}       &0.33$^a$&{-}  &{b} &1.8$^a$&{-}  &14    &6.60&{-}  &51  &14.79    &0.12&75 \\
				7 &{-}&{-}&{-}       &{-}     &{-}  &{-} &0.74$^{a}$&{-}&{b}  &3.0$^a$&{-}&14 &9.63     &0.10&43 \\
				8 &{-}&{-}&{-}       &{-}     &{-}  &{-} &{-}&{-}&{-}         &1.0$^a$&{-}&{b}&5.2$^a$  &{-}  &{b}  \\
			\end{tabular}
				\footnotetext[1]{Parameters with label a were kept fixed during final fit.}
			\footnotetext[2]{The label b indicates that the crossing is beyond our data range or not existing (compare Figure \ref{fig:lica_Energien}).}
		\end{minipage}
	\end{table*}
	
For the simulation of the spectra as shown in Fig.\ref{fig:lica_sim} b) we need the potential of the ground state. This potential has been significantly improved compared to Ref.  \cite{ivanova_x_2011}  by enlarging the range of observed rotational levels up to N=123. The new potential  is given in Table \ref{pot_X} together with the spin-rotation parameters for which we experimentally confirmed the sign by the evaluation of the perturbation. 
\begin{table}[h]
\fontsize{8pt}{8pt}\selectfont 
\caption{Parameters of the
analytic representation of the \Xstate\ state potential of LiCa and spin-rotation parameters for $^7$Li$^{40}$Ca. The energy reference is the minimum of the potential. Parameters with an asterisk $^\ast$ ensure smooth extrapolation of the potential at $R_{in}$.\label{pot_X}} 
\vspace{2mm}
\begin{tabular*}{0.8\linewidth}{@{\extracolsep{\fill}}ccc}
\hline
\multicolumn{3}{c}{$R < R_\mathrm{in}= 2.753$ \AA} \\
\hline
$A^\ast$ & $-0.22217777\times 10^{4}$ &\wn \\
$B^\ast$ & $ 0.19891763\times 10^{7}$ &\wn \AA $^{6}$ \\
\hline
\multicolumn{3}{c}{$R_\mathrm{in} \leq R \leq R_\mathrm{out}=  7.410$ \AA} \\
\hline
$b$ & $-0.150$ \\
$R_\mathrm{m}$ & $3.3558559$ &\AA \\
 $a_{0}$ & $ 0.0000$ &\wn\\
 $a_{1}$ & $ 1.28299104401761443$ &\wn\\
 $a_{2}$ & $ 0.294670379190349231\times 10^{ 5}$& \wn\\
 $a_{3}$ & $-0.252704126684650764\times 10^{ 5}$& \wn\\
 $a_{4}$ & $-0.585021073574163092\times 10^{ 5}$& \wn\\
 $a_{5}$ & $-0.271380018049823448\times 10^{ 5}$& \wn\\
 $a_{6}$ & $ 0.630225745352516242\times 10^{ 5}$& \wn\\
 $a_{7}$ & $ 0.335791425026657758\times 10^{ 6}$& \wn\\
 $a_{8}$ & $ 0.923816940807595733\times 10^{ 6}$& \wn\\
 $a_{9}$ & $-0.952709968557935557\times 10^{ 6}$& \wn\\
$a_{10}$ & $-0.916034917693250440\times 10^{ 7}$& \wn\\
$a_{11}$ & $ 0.607979993761450518\times 10^{ 7}$& \wn\\
$a_{12}$ & $ 0.411707171526412815\times 10^{ 8}$& \wn\\
$a_{13}$ & $-0.732652303029812723\times 10^{ 8}$& \wn\\
$a_{14}$ & $ 0.360209838290596455\times 10^{ 8}$& \wn\\
\hline
\multicolumn{3}{c}{$R_\mathrm{out} < R$}\\
\hline
${U_\infty}$ & 2605.8946 &\wn \\
${C_6}$      & 0.820 $\times 10^{7}$ &\wn\AA$^6$ \\
${C_{8}}$    & 0.217122200$\times 10^{9}$ &\wn\AA$^8$ \\
${C_{10}}$   & 0.225401168$\times 10^{10}$&\wn\AA$^{10}$ \\
\hline
\multicolumn{3}{c}{spin-rotation in Dunham notation}\\
\hline
$\gamma_{00}$ & 0.0033113& \wn\\
$\gamma_{10}$ &-0.00021387& \wn\\
\hline
\hline
\multicolumn{3}{c}{Derived constants:} \\
\hline
\multicolumn{3}{l}{equilibrium separation:\hspace{0.4cm} $R_e^X$= 3.35579(10) \AA} \\
\multicolumn{3}{l}{dissociation energy:\hspace{0.7cm} $D_e^X$= 2605.9(100) \wn}\\  %
\hline
\end{tabular*}
\end{table}

The energy reference in this paper is the minimum of the ground state potential as it is convention in molecular physics. But this definition depends on the mathematical representation of the potential. Using other forms will lead to different extrapolations to the minimum. To easily transform the energy reference used here to other forms, we calculated the energy position of the quantum state N=0, F$_1$, v=0 and obtained for $^7$Li$^{40}$Ca 100.1267 \wn. This quantum state can be calculated with other potential representations in their energy reference and provides the connection between the different models.

\end{document}